# PION CONDENSATION DURING THE HADRONIZATION OF THE QUARK-GLUON PLASMA IN ULTRA-RELATIVISTIC HEAVY-ION COLLISIONS [1]

W. Florkowski [2,3] and M. Abu-Samreh [4]

Institut für Theoretische Physik, Universität Heidelberg
D-69120 Heidelberg, Philosophenweg 19, Germany

**Abstract:** Using a simple kinetic model we study whether a pion condensate can be formed during the hadronization of the quark-gluon plasma in ultra-relativistic heavy-ion collisions. If hadronization proceeds fast and the pion effective mass is close to zero, some pions condense into the zero momentum state. On the other hand, for the pion effective mass equal to the vacuum one or larger ($m_\pi \geq 140$ MeV) no onset of the Bose-Einstein condensation is observed. The constraints on the condensation process coming from the requirement of the entropy increase are discussed in detail.

HEIDELBERG, FEBRUARY 1995

---

[1] Research supported in part by the Federal Minister for Research and Technology (BMFT) grant number 06 HD 742 (0), and the Deutsche Forschungsgemeinschaft, grant number Hu 233/4-3

[2] Permanent address: H. Niewodniczański Institute of Nuclear Physics, ul. Radzikowskiego 152, PL-31-342 Kraków, Poland.

[3] E-mail address: wojtek@hobbit.mpi-hd.mpg.de.

[4] Permanent address: College of Science and Technology, Jerusalem Abu-Deis, P.O. Box 20002, West Bank, Israel.

# 1. Introduction

In ultra-relativistic heavy-ion collisions [1, 2] large numbers of pions can be created in small space-time regions. In such a situation we may ask the question: What is the probability that a part of them will form a condensate? In the recent paper of Greiner, Gong and Müller [3], it has been pointed out that the occurrence of condensed pions can only take place if the system evolves far away from thermal equilibrium. They argue that if the hadronization of a deconfined quark-gluon plasma (QGP) occurs rather slowly and the initial pion temperature is of the order of 170 MeV, then the inelastic reactions changing the pion number enforce the pion chemical potential to be zero. The detailed study [3] of the subsequent adiabatic expansion of such a pion gas indicates that the pion chemical potential builds up. However, it remains always below the critical value, independent of the interactions assumed. Consequently, the pion gas in thermal equilibrium can never reach the onset of Bose-Einstein condensation (BEC).

Having the above results in mind, in the present paper we are going to consider a non-equilibrium hadronization of the QGP and to study whether *already* during such a process a part of pions can produce a condensate. To some extent, our investigations can be regarded as supplementary to those of [3] since we concentrate on possible pion condensation occurring at earlier times and employ a different approach, i.e., we solve the system of kinetic equations in a model study, that allows us to study the non-equilibrium processes.

In our calculation, we assume that the hadronization is fast, so, at least for the initial stages of the evolution of the system, we can assume that it happens within a constant volume. Nevertheless, the produced pions can leave this volume (evaporate), which is crucial for the overall increase of the entropy during this reaction. One can check that the *total* hadronization of the QGP into the pion gas taking place in a constant volume and at constant energy is impossible, since, roughly speaking, it leads to the reduction of the number of degrees of freedom (if one takes into account only pions and neglects resonances). On the other hand a *partial* hadronization is possible, and, if connected with the simultaneous evaporation of mesons, can lead in a sequence of steps to the complete hadronization of the plasma and to the increase of entropy.

We also assume that the reactions conserving the pion number dominate, i.e., except for the very process of the hadronization, the other reactions do not change the pion number. Such an approach can be justified if the cross sections for the inelastic, pion number changing reactions such as $\pi\pi\pi \to \pi\pi$ are much smaller than those for elastic ones. Since the in-medium cross sections are rather poorly known, we think that it is useful to look for the consequences of this hypothesis. In such a case, a very dense system of pions is created just at the beginning of the hadronization process, and, perhaps, a part of them will form a condensate.

In order to study quantitatively the possibility of appearance of the pion condensate, we shall solve a system of kinetic equations describing the transformation of quarks and gluons into pions. To facilitate these considerations, we assume that the phase space distribution functions are functions of momentum $p = |\mathbf{p}|$ and time $t$ only. Thus the particles are always uniformly distributed in space and the system is isotropic. Moreover, we consider the kinetic equations in the relaxation time approximation, reducing the problem to solving a system of intego-differential equations.

A novel feature of our approach is the incorporation of the phenomenon of the BEC into such a relaxation time formalism. To some extent, our description is close to the statistical one: For a given pion distribution function $f_\pi(p,t)$, we check whether the corresponding thermal distribution (i.e., the distribution which has the thermal shape and gives the same energy and particle densities as the function $f_\pi(p,t)$) contains a condensate part. If so, then the pion collisions that always bring the distribution function $f_\pi(p,t)$ to the thermal one with the speed characterized by the relaxation time, will try to force some part of pions to form the condensate.

Due to the pion interactions between themselves and with quarks and gluons, the pion dispersion relation in medium can change. In order to take into account some aspects of this problem, we shall consider different pion masses in our calculations, i.e., we shall take the pion dispersion relation of the (free) form $E_p^2 = m_\pi^2 + p^2$ with different values of $m_\pi$. In particular, we shall consider the cases $m_\pi = 0$ and $m_\pi = 140$ MeV.

The paper is organized as follows: In the next Section, we introduce the kinetic equations in the relaxation time approximation, define the thermal functions, discuss the increase of entropy, and calculate the relaxation times. Section 3 contains our results. We summarize and conclude in Section 4.

## 2. Kinetic description of the hadronization of the QGP

*i) Kinetic equations in the relaxation time approximation*

We assume that the hadronization reactions take place in constant volume $V_{pl}$ and that the system is spherically symmetric, i.e., $V_{pl} = \frac{4}{3}\pi R_{pl}^3$ where $R_{pl}$ is the radius of the fireball containing quarks, gluons and pions (quarks and gluons cannot escape from this mixed phase region). The number of pions evaporated per unit time can be estimated to be $\frac{1}{2} f_\pi(p,t) S_{pl} \overline{v}_p$, where $S_{pl}$ is the area of the fireball surface and $\overline{v}_p = 2p/\pi\sqrt{p^2 + m_\pi^2}$ is the average pion velocity perpendicular to the surface. The factor $\frac{1}{2}$ accounts for the fact that only one half of the particles situated close to the surface and having momentum $p = |\mathbf{p}|$ move *outwards* and can, in consequence, leave the system. The volume occupied by the emitted pions is defined as $V_{had}(t) = \frac{4}{3}\pi \left[ R_{had}^3(t) - R_{pl}^3 \right]$ where $R_{had}(t)$ is the radius of our total hadronic system. We assume that $R_{had}(t)$ increases in time with the velocity of light.

Having this geometry in mind we study the following set of the kinetic equations (they are a generalization of the equations introduced in [4] by including gluons and, as we shall see later, by considering the quantum statistics)

$$\frac{df_q(p,t)}{dt} = -\frac{f_q(p,t) - f_{q,th}(p,t)}{\tau_{th}(t)} - \frac{f_q(p,t)}{\tau_{q,had}(t)}, \tag{1}$$

$$\frac{df_g(p,t)}{dt} = -\frac{f_g(p,t) - f_{g,th}(p,t)}{\tau_{th}(t)} - \frac{f_g(p,t)}{\tau_{g,had}(t)}, \tag{2}$$

$$\frac{df_\pi(p,t)}{dt} = -\frac{f_\pi(p,t) - f_{\pi,th}(p,t)}{\tau_{th}(t)} + \frac{f_q(p,t)}{\tau_{q,had}(t)} R_q + \frac{f_g(p,t)}{\tau_{g,had}(t)} R_g - f_\pi(p,t) \frac{S_{pl}}{2V_{pl}} \overline{v}_p, \tag{3}$$

$$\frac{df_{em}(p,t)}{dt} = -\frac{f_{em}(p,t)}{V_{had}(t)} \frac{dV_{had}(t)}{dt} + f_\pi(p,t) \frac{S_{pl}}{2V_{had}(t)} \overline{v}_p. \tag{4}$$

Eq. (1) determines the time evolution of the quark distribution function $f_q(p,t)$ (it describes both quarks and antiquarks, i.e., $f_q(p,t) = f_{quarks}(p,t) + f_{antiquarks}(p,t)$). The first term on the right-hand-side (RHS) of Eq. (1) is the collision term written in the relaxation time approximation; it is responsible for the thermalization of quarks since the distribution function $f_q(p,t)$ is always attracted to the thermal one $f_{q,th}(p,t)$. The second term describes the loss of quarks due to the hadronization process. We note that the number of quarks in the system will be always equal to the number of antiquarks. Consequently, we study the hadronization of the baryon-free plasma and this might lead to the creation of the condensate which carries *no isospin*.

Eq. (2) has the form completely analogous to that of Eq. (1) and describes the thermalization and hadronization of gluons. In Eq. (3), for the time evolution of the distribution functions of pions, the first three terms on the RHS describe pion thermalization and pion creation due to the hadronization of quarks and gluons, respectively. The factors $R_q = \sqrt{(p^2 + m_q^2)/(p^2 + m_\pi^2)}$ and $R_g = \sqrt{(p^2 + m_g^2)/(p^2 + m_\pi^2)}$ (here $m_q, m_g$ and $m_\pi$ is the quark, gluon and the pion mass) are included in order to have the overall conservation of the energy in the system. The fourth term accounts for the evaporation of pions.

Our last kinetic equation (4) describes the time changes of the distribution function of the pions which have been emitted from the reaction volume. The first term on the RHS of Eq. (4) appears due to the expansion of these pions into the vacuum, and the second term collects the pions acting as a sink in Eq. (3).

The effects of Bose-Einstein condensation can be included into our description if we represent the pion distribution function as a sum of two terms, namely

$$f_\pi(p,t) = f_\pi^>(p,t) + c_\pi(t)(2\pi)^3 \delta^{(3)}(p). \tag{5}$$

The analogous decomposition holds also for the function $f_{\pi,th}(p,t)$. In this case Eq. (3) can be rewritten as follows

$$\frac{df_\pi^>(p,t)}{dt} = -\frac{f_\pi^>(p,t) - f_{\pi,th}^>(p,t)}{\tau_{th}(t)} + \frac{f_q(p,t)}{\tau_{q,had}(t)} R_q + \frac{f_g(p,t)}{\tau_{g,had}(t)} R_g - f_\pi^>(p,t)\frac{S_{pl}}{2V_{pl}}\overline{v}_p, \tag{6}$$

$$\frac{dc_\pi(t)}{dt} = -\frac{c_\pi(t) - c_{\pi,th}(t)}{\tau_{th}(t)}, \tag{7}$$

where we have explicitely separated the smooth and the singular parts of the pion distributions.

*ii) Thermal distributions*

Eqs. (1) - (4) can be solved for any initial conditions (i.e., assuming some particular form of the distribution functions at the initial time $t = 0$) provided the thermal distribution functions $f_{q,th}(p,t)$, $f_{g,th}(p,t)$ and $f_{\pi,th}(p,t)$ are known at all times. Since the first terms on the RHS of Eqs. (1), (2) and (3) represent the collision terms (in the relaxation time approximation) they must obey the symmetry leading to the energy conservation. This gives the following constraint for $f_{q,th}(p,t)$, $f_{g,th}(p,t)$ and $f_{\pi,th}(p,t)$ at each time $t$

$$\int \frac{d^3p}{(2\pi)^3} \sqrt{p^2 + m_i^2} \left[f_i(p,t) - f_{i,th}(p,t)\right] = 0, \qquad (8)$$

where $i = q, g$ or $\pi$. Our collision terms for quarks and gluons account for elastic and inelastic scattering, which leads to the change in the number of participating quarks and gluons. On the other hand, in the case of pions, we assume that their number is not changed by collisions (*except for the hadronization processes, but these are already included in the separate terms*). Consequently, the pion collision term should posess the additional symmetry, namely

$$\int \frac{d^3p}{(2\pi)^3} \left[f_\pi(p,t) - f_{\pi,th}(p,t)\right] = 0. \qquad (9)$$

Eqs. (8) and (9) can be used to determine the thermodynamic parameters appearing in the thermal distributions. For quarks and gluons the latter have the form

$$f_{q,th}(p,t) = g_q \left\{ \exp\left[-\frac{\sqrt{p^2 + m_q^2}}{T_q(t)}\right] + 1 \right\}^{-1} \qquad (10)$$

and

$$f_{g,th}(p,t) = g_g \left\{ \exp\left[-\frac{p}{T_g(t)}\right] - 1 \right\}^{-1}, \qquad (11)$$

where we have introduced the degeneracy factors: $g_q = 24$ (quarks and antiquarks having two different spin projections, 3 colors and 2 flavors) and $g_g = 16$ (two spin projections times 8 colors). In the case of pions the thermal background distribution function can be written as

$$f_{\pi,th}(p,t) = g_\pi \left\{ \exp\left[-\frac{\sqrt{p^2 + m_\pi^2} - \mu_\pi(t)}{T_\pi(t)}\right] - 1 \right\}^{-1} + c_{\pi,th}(t)\,(2\pi)^3 \delta^{(3)}(p), \qquad (12)$$

where $g_\pi = 3$ (three different values of the isospin). For pions we have generalized the thermal distribution in order to include also the effect of the condensation. The number of pions in the thermal condensate equals $c_{\pi,th}(t)$, and $\mu_\pi(t)$ is the pion chemical potential. The quantities $T_i(t)$ appearing in Eqs. (10) - (12) are the quark, gluon and pion temperatures, respectively.

In the situation without the pion condensation Eqs. (8) and (9) can be fulfilled by finding the appropriate values of $T_i(t)$ and $\mu_\pi(t)$, and by setting $c_\pi(t) = c_{\pi,th}(t) = 0$. For increasing pion densities the pion chemical potential $\mu_\pi(t)$ increases and at some stage (let us say for $t = t_{cond}$) it may happen that it reaches its maximal value, i.e., it becomes equal to the pion effective mass $m_\pi$. This is exactly the moment when the pion condensate sets in. For larger pion densities we have to assume that a part of them occupies macroscopically the zero momentum state. In this situation Eqs. (8) and (9), considered for $i = \pi$, are two equations determining the pion temperature $T_\pi(t)$ and the pion thermal condensate $c_{\pi,th}(t)$ (the pion chemical potential for $t \geq t_{cond}$ becomes constant and takes on its critical value $\mu_\pi = m_\pi$). If $c_{\pi,th}(t)$ starts out as being different from zero, then according to our Eq. (7), the actual distribution function $f_\pi(p,t)$ also develops a singular part $c_\pi(t)$.

*iii) Energy conservation law and the increase of entropy*

Our kinetic equations (1) - (4) are constructed in such a way that the total energy $E(t)$ of quarks, gluons and pions (in and out of the fireball) is conserved. Defining the particle energy densities as

$$\varepsilon_i(t) = \int \frac{d^3p}{(2\pi)^3} \sqrt{p^2 + m_i^2} f_i(p,t), \tag{13}$$

we can explicitly write this conservation law in the form

$$\frac{d}{dt}E(t) = \frac{d}{dt}\left\{V_{pl}\left[\varepsilon_q(t) + \varepsilon_g(t) + \varepsilon_\pi(t)\right] + V_{had}(t)\varepsilon_{em}(t)\right\} = 0. \tag{14}$$

The total entropy of our mixture can be defined as follows

$$S(t) = V_{pl}\left[s_q(t) + s_g(t) + s_\pi(t)\right] + V_{had}(t)s_{em}(t), \tag{15}$$

where [5]

$$s_i(t) = -g_i \int \frac{d^3p}{(2\pi)^3} \left\{ \frac{f_i(p,t)}{g_i} \ln \frac{f_i(p,t)}{g_i} - \frac{1}{\epsilon_i}\left[1 + \epsilon_i \frac{f_i(p,t)}{g_i}\right] \ln \left[1 + \epsilon_i \frac{f_i(p,t)}{g_i}\right] \right\}. \tag{16}$$

Here the quantity $\epsilon_i$ equals 1 for gluons and pions, and $-1$ for quarks (the classical Boltzmann definition corresponds to the limit $\epsilon \to 0$). It is important to notice that in the case of pions the distribution function $f_\pi(p,t)$ in (16) should correspond to $f_\pi^>(p,t)$ since the condensate does not contribute to the entropy of the system.

Although Eqs. (1) - (4) lead to the conservation law (14) they do not guarantee, *in general*, that their solutions correspond to states of increasing entropy. Consequently, for each solution we check the behaviour of the function $S(t)$, and accept as the physical ones only these for which $dS(t)/dt \geq 0$ (for each $t$). This fact indicates some difficiency of the present description of hadronization, namely it cannot be valid for arbitrary initial conditions or parameters. One might expect this type of difficulty, since some physical processes have been neglected in our approach and they might become important for a given class of initial conditions. In particular, we have neglected processes such as: (i) an expansion of the volume of hadronizing matter, (ii) the possible back reactions of pions into quarks and gluons, (iii) the production of pion resonances, or (iv) the process in which a quark tries to leave the plasma, stretches the string which later decays producing the meson outside. One observes that all these processes decrease chance of producing of a condensate.

The study of the time evolution of the entropy shows, in particular, that it is crucial to take into account the appropriate balance between the hadronization and evaporation rates, i.e., the hadronization cannot be too fast with respect to evaporation. In our case, we control the speed of hadronization by considering different values of the hadronization cross section $\sigma$, which is regarded as a parameter. The evaporation rate is determined mainly by the geometry of the system, in particular, by the surface to volume ratio. As a consequence, we can always study *optimal* conditions for creation of the condensate (i.e., for a given geometry we consider the largest value of $\sigma$ which allows for entropy growth) and check whether it has been *really* produced.

*iv) Relaxation times and cross sections*

Let us now turn to the discussion of the relaxation times appearing in our kinetic equations (1) - (4). The analysis of the exact collision term in the Boltzmann kinetic equation (analogous to that in [6]) leads to an expression for the average time for *hadronization* of two quarks into two pions

$$\frac{1}{\tau_{q,had}(p,t)} = \frac{1}{2\sqrt{p^2+m_q^2}} \int \frac{d^3p_1}{(2\pi)^3\sqrt{p_1^2+m_q^2}} f_q(p_1,t) F_{qq}(s) \sigma_{q\bar{q}\to\pi\pi}(s), \qquad (17)$$

where $\sigma_{q\bar{q}\to\pi\pi}(s)$ denotes the total cross section for this process. The relativistic flux factor of incoming quarks is $F_{qq}(s) = \frac{1}{2}\sqrt{s(s-4m_q^2)}$, with $\sqrt{s}$ being the center-of-mass energy of the quarks with momenta $p$ and $p_1$, respectively. The analogous expression holds, of course, for the gluon hadronization time $\tau_{g,had}(p,t)$.

Here we also make simplifying assumptions: In the limit of vanishing masses (this case is exact for gluons and represents a very good approximation for current quarks) and for the constant cross sections ($\sigma$'s independent of $s$) we find

$$\frac{1}{\tau_{q,had}(t)} = \frac{1}{2}\, n_q(t)\, \sigma_{q\bar{q}\to\pi\pi}\,, \quad \frac{1}{\tau_{g,had}(t)} = \frac{1}{2}\, n_g(t)\, \sigma_{gg\to\pi\pi}\,, \tag{18}$$

where

$$n_i(t) = \int \frac{d^3p}{(2\pi)^3} f_i(p,t) \tag{19}$$

are particle's number densities. Consequently, one finds that the hadronization relaxation times are functions of time only. We note that if the cross sections depend on $s$, the relaxation times (even for $m_i = 0$) can exhibit a $p$-dependence. In particular, the singularities of the cross sections can be reflected in the singularities of the inverse relaxation times. For the discussion of this point in the context of the critical scattering we refer the reader to Ref. [4].

The relaxation time for the *thermalization* of quarks (determined mainly by the elastic collisions) can be written in the form

$$\begin{aligned}
\frac{1}{\tau_{th}(p,t)} = &\ \frac{1}{2\sqrt{p^2+m_q^2}} \int \frac{d^3p_1}{(2\pi)^3\sqrt{p_1^2+m_q^2}} f_q(p_1,t) F_{qq}(s) \sigma_{qq\to qq}(s) \\
&+ \frac{1}{2p} \int \frac{d^3p_1}{(2\pi)^3 p_1} f_g(p_1,t) F_{qg}(s) \sigma_{qg\to qg}(s) \\
&+ \frac{1}{2\sqrt{p^2+m_\pi^2}} \int \frac{d^3p_1}{(2\pi)^3\sqrt{p_1^2+m_\pi^2}} f_\pi(p_1,t) F_{q\pi}(s) \sigma_{q\pi\to q\pi}(s),
\end{aligned} \tag{20}$$

where the three terms on the RHS of Eq. (20) appear due to the quark-quark, quark-gluon and quark-pion collisions, respectively. The formulae for the thermalization times of gluons and pions can be obtained from Eq. (20) by appropriate changes of the indices $q, g$ and $\pi$. Considering again the massless limit and assuming that the cross sections are independent of $s$, we find

$$\frac{1}{\tau_{th}(t)} = \frac{1}{2} \left[ n_q(t) \sigma_{qq\to qq} + n_g(t) \sigma_{qg\to qg} + n_\pi(t) \sigma_{q\pi\to q\pi} \right]. \tag{21}$$

In order to reduce the number of independent parameters we shall set all the cross sections appearing in Eqs. (17) and (20) to be the same and equal $\sigma$. In this case the thermalization time is the same for quarks, gluons and pions.

We note that in our calculations we shall consider different effective pion masses, in particular, we shall take into account the normal vacuum mass $m_\pi = 140$ MeV. In such a case the approximation $m_\pi \approx 0$ is not as good as the approximation $m_q \approx 0$. Nevertheless, we have checked that the inclusion of the finite meson mass here (i.e., in the calculation of the relaxation times) represents only a small correction. Much more important are effects coming from the appearance of the finite and large (compatible with the temperature) pion mass in the thermal functions (12) since in this case $m_\pi$ is the argument of the exponent; these corrections will be exactly taken into account.

## 3. Results

The initial conditions assumed for solving the kinetic equations (1) - (4) correspond to a quark-gluon plasma in thermal and chemical equilibrium, which is confined in a spherical fireball. This means that the quark and gluon distribution functions have initially the thermal form (10) and (11). Moreover, at the beginning of the process there are no pions in the system, i.e., $f_\pi(p, t = 0) = 0$.

For $t = 0$ we specify the initial plasma temperature $T_0$ and the plasma radius $R_{pl}$ (initially $V_{had} = 0$). The additional parameters are: the value of the hadronization/thermalization cross section $\sigma$, and the quark and pion masses (the gluon mass is always taken to be zero). We do the calculations for different values of the pion mass since we want to study the sensitivity of our results to the change of the effective pion mass resulting from the pion interactions with other particles in the medium.

Let us first consider the case when all the masses are equal to zero ($m_q = m_g = m_\pi = 0$). In Figs. 2 and 3 we show our results for the initial temperature $T_0 = 150$ MeV, $\sigma = 9$ mb and $R_{pl} = 2$ fm ($V_{pl} \approx 34$ fm$^3$). This situation corresponds to the initial energy density of 0.8 GeV/fm$^3$, and the total number of particles (quarks and gluons) is 59. The initial total entropy calculated from (15) is 227. The choice $T_0 = 150$ MeV is motivated by the results of the lattice simulations of QCD, which indicate such a temperature of the phase transition.

In Fig. 1 we plot the time evolution of the thermodynamic parameters characterizing the pion thermal distribution. The upper curve describes the changes of the temperature and the lower one shows the behaviour of the chemical potential. The pion temperature first increases and later saturates at 200 MeV. In our opinion, the increase of the pion temperature is due to the fact that we do not have pion number changing processes. (At the same time the quark and gluon temperatures remain approximately constant.) The pion chemical potential is initially very large and negative since the pion density is very low, however later it builds up very fast and for $t = 0.4$ fm it already reaches its critical value, i.e., in this case zero.

In Fig. 2 the time evolution of the particle's numbers has been drawn. We can see that the numbers of quarks and gluons decrease in time. Initially the number of pions increases rather fast since the process of evaporation (proportional to $f_\pi(p,t)$) is rather weak. At the moment when the pion chemical potential takes on its critical value, some fraction of them starts to occupy the zero momentum state, i.e., for $t \geq 0.4$ fm we find $N_{cond} \geq 0$. For larger times the number of pions in the fireball saturates, however the number of evaporated pions starts to increase rapidly (at this stage the evaporation becomes a very efficient process). After 2 fm the average number of pions in the condensate is 2.

In the considered case (corresponding to Figs. 1 and 2) the entropy increases only by 7%. In fact, if we did the calculations for larger values of the hadronization cross sections, we would find that the entropy derivative becomes negative in some time intervals. In this place we are touching the problems already discussed before: For some initial conditions (or values of the parameters) one should include more processes into the kinetic description. Nevertheless, we expect that such improvements can have only little effect on general conditions concerning the condensate creation. Our specific calculation shows that for very large $\sigma$ (larger chances for condensate cration) we have to take into account other processes (reducing the probability of pion condensation). Consequently, the overall result of more complicated calculations can be very similar to the present ones.

Let us now discuss the results obtained for the following set of masses: $m_q = 10$ MeV, $m_g = 0$, and $m_\pi = 140$ MeV. The initial temperature is again equal to 150 MeV and $R_{pl} = 2$ fm. The hadronization cross section is 7 mb since for larger values the total entropy would be (in some time intervals) a decreasing function of time. One can notice that in this case (massive pions) the hadronization process cannot be so fast as in the previous case since the evaporation is slower. In Fig. 3 (analogous to Fig. 1) we show the time evolution of the thermodynamic parameters: The pion temperature slightly increases, whereas the pion chemical potential grows up substantially. Nevertheless, its value saturates for larger times and remains always below the critical one (in this case equal to 140 MeV). Consequently, in this case pions do not condense.

## 4. Conclusions

In this paper we have studied the hadronization of a QGP into the pion gas and investigated the possibility of creation of the pion condensate during such a process. For fixed initial temperature we find that the results depend on the pion effective mass in medium.

For massless pions, a small fraction of them will condense. On the other hand for pion effective masses compatible with its vacuum mass, no onset of Bose-Einstein condensation is found. The requirement that the entropy should increase during the hadronization leads to the strong constraints on the possibility of creation of a very dense pion system and, in consequence, on possibility of pion condensation.

We studied the initial stages of hadronization since for larger times other processes should become important (e.g., pion number changing reactions or volume expansion) which has been neglected in our approach. Such processes will diminish the chances of producing a condensate and, consequently, if it is not produced fast then it is probably not produced at all. Another open question is whether a condensate once produced has chances to survive. In our case we observe the formation of the condensate for massless pions. The effect is not large and, perhaps, can be destroyed when the particles gain their true vacuum masses.

*Acknowledgements:* We thank Jörg Hüfner for the critical reading of the manuscript, continuous encouragement, and very warm hospitality at the Heidelberg University. We also thank Sandi Klevansky for critical comments concerning the manuscript. One of us (M. A.-S.) is grateful to the German Exchange Service (DAAD) for making his stay in Heidelberg possible.

**Figure Caption**

**Fig. 1** Pion temperature and pion chemical potential as the functions of time ($T_0 = 150$ MeV, $\sigma = 9$ mb, $m_q = m_g = m_\pi = 0$, and $R_{pl} = 2$ fm).

**Fig. 2** Time evolution of the particle's numbers. The parameters as in Fig. 1.

**Fig. 3** Pion temperature and pion chemical potential as the functions of time ($T_0 = 150$ MeV, $\sigma = 7$ mb, $m_q = 10$ MeV, $m_g = 0$, $m_\pi = 140$ MeV, and $R_{pl} = 2$ fm).

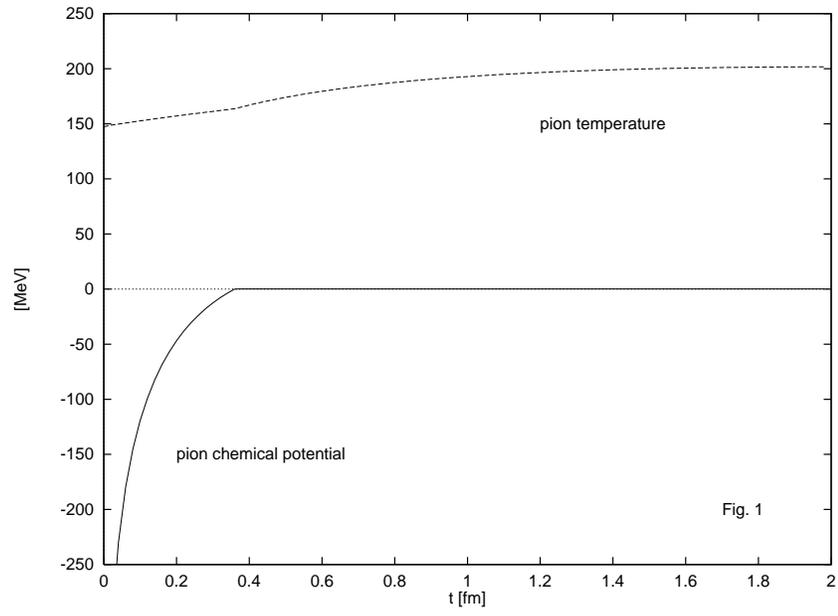

Fig. 1

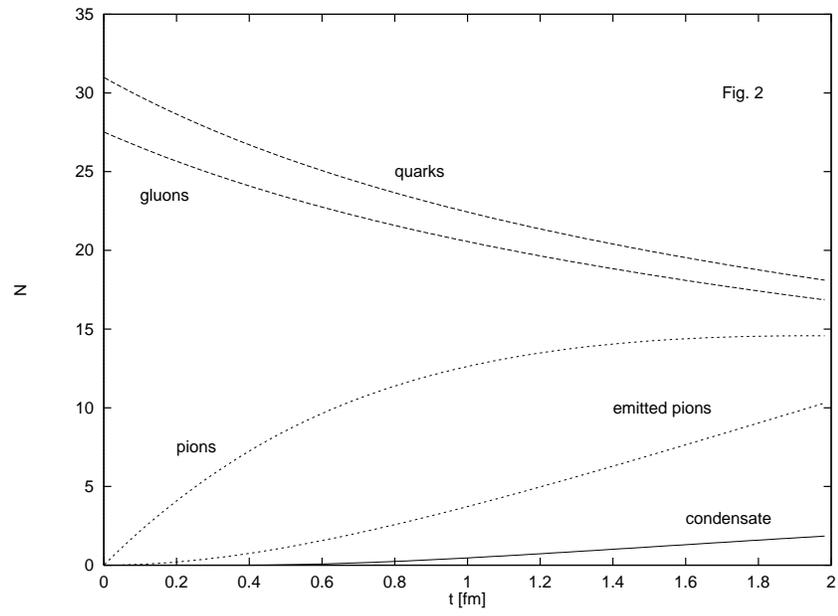

Fig. 2

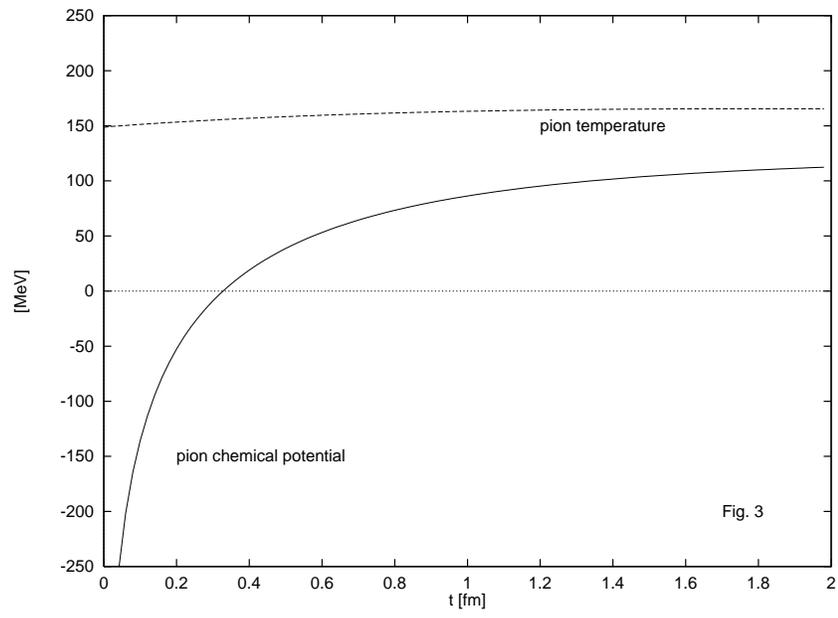

Fig. 3